\documentclass[twocolumn,showpacs,preprintnumbers,amsmath,amssymb,
floatfix,nofootinbib]{revtex4}
\usepackage{graphicx}

\newcommand{\comment}[1]{}
\newcommand{\eq}[2]{\begin{equation}\label{#1} #2\end{equation}}

\newcommand{\nco}{\newcommand}
\nco{\beq}{\begin{equation}} \nco{\eeq}{\end{equation}}
\nco{\beqa}{\begin{eqnarray}} \nco{\eeqa}{\end{eqnarray}}

\nco{\sss}{\scriptscriptstyle} \nco{\dphi}{\varphi}
\nco{\lsim}{\lesssim}
\nco{\gsim}{\mbox{\raisebox{-.6ex}{~$\stackrel{>}{\sim}$~}}}

\begin{document}

\title{Predictions of the causal entropic principle\\
for environmental conditions of the universe}

\author{James M. Cline}\email{jcline@physics.mcgill.ca}
\author{Andrew R. Frey}\email{frey@hep.physics.mcgill.ca}
\author{Gilbert Holder}\email{holder@hep.physics.mcgill.ca}

\affiliation{%
Physics Department, McGill University,
3600 University Street, Montr\'eal, Qu\'ebec, Canada H3A 2T8
}

\date{August 2007}

\begin{abstract} 

The causal entropic principle has been proposed as a superior
alternative to the anthropic principle for understanding the
magnitude of the cosmological constant.  In this approach, the
probability to create observers is assumed to be proportional to the
entropy production  $\Delta S$ in a maximal causally connected
region --- the causal diamond.  We improve on the original treatment
by better quantifying the entropy production due to stars, using an
analytic model for the star formation history which accurately 
accounts for changes in cosmological parameters.  We calculate the
dependence of $\Delta S$ on the density contrast $Q=\delta\rho/\rho$,
and find that our universe is much closer to the most probable value
of $Q$ than in the usual anthropic approach and that probabilities
are relatively weakly dependent on this amplitude.  In addition, we make first
estimates of the dependence of $\Delta S$ on the baryon fraction and 
overall matter abundance.  Finally, we also explore the
possibility that decays of dark matter, suggested by various observed
gamma ray excesses, might produce a comparable amount of entropy to
stars.

\end{abstract}

\maketitle

\section{Introduction}

The nature of the cosmological constant problem has changed since it
was fully recognized in the 1980s.  At that time, a seemingly
natural solution was to find a mechanism which would explain why
$\Lambda$ should be zero.  Among the most promising such ideas were
those of Hawking \cite{Hawking}, elaborated by Coleman
\cite{Coleman}, in which the Euclidean path integral for gravity was
interpreted as a wave function for the universe, shown to be strongly
peaked at $\Lambda = 0$.  However we now know with a high
degree of confidence that $\Lambda$ is not zero, due to observations
of distant type Ia supernovae \cite{Riess} combined with constraints
on the flatness of the universe from the CMB \cite{Spergel} and
complementary information from the x-ray baryon fraction in galactic
clusters \cite{Allen}.  

Two notable ideas from the 1980's predicted that $\Lambda$ should be
nonzero: the membrane creation mechanism of Brown and Teitelboim
(BT) \cite{BT} (building on work in \cite{linde}),  
involving tunneling between false vacua with different
values of $\Lambda$, and the anthropic argument of  Weinberg
\cite{Weinberg}, which predicted that 
\beq	
-\Lambda_{\mathrm{obs}} \lsim \Lambda \lsim 100\, \Lambda_{\mathrm{obs}}
\label{abound}
\eeq
based on the requirement that galaxies should be able to form and
thus give rise to observers before the universe collapsed or 
expanded too quickly.  These approaches have recently gained support 
in a number of ways from the string theory community: by the explicit
realization of the BT mechanism within heterotic M-theory and
type II string theory \cite{BP}
and the realization that the vacuum structure of string theory is
a vast landscape \cite{Susskind} encompassing more than $10^{500}$
possible values of $\Lambda$ \cite{Douglas}.  The picture that has
emerged is that this multitude of vacua can be simultaneously realized
through the process of eternal inflation \cite{Linde}.  Tunneling
\textit{\`a la} BT from the eternally inflating regions 
would populate universes like ours with small
$\Lambda$, as well as those with much larger $\Lambda$.  The 
latter would be incompatible with life, and we
should disregard them if we are interested in explaining properties of
universes which can admit the existence of observers.  This is the
weakest form of the anthropic principle; it does not insist that
physics must be compatible with our existence, only that we take our
existence as a data point in putting experimental constraints on the
landscape of vacua.

This point of view has been strongly criticized on the grounds that
it admits no real predictions, only postdictions.  Moreover, the
anthropic explanation of $\Lambda$ has been weakened by the 
discovery of dwarf galaxies at $z\sim 10$, proving that structure
formed earlier than $z\sim 4$ as was supposed in 
\cite{Weinberg}. Since the bound (\ref{abound}) scales like
$(1+z)^3$, the anthropic constraint is weakened by an order of
magnitude. To counteract this, it has been argued that only galaxies
of a certain minimum size are capable of retaining the heavy elements
needed to sustain life \cite{Tegmark}, 
but such assumptions seem questionable and make the
anthropic principle appear to be increasingly arbitrary. The
anthropic bound on $\Lambda$ is further weakened by allowing other
cosmological parameters to vary at the same time.  For example, the
primordial density contrast  $Q=\delta\rho/\rho\sim 2\times 10^{-5}$
in our universe, but, if $Q$ were larger, structure could form earlier
despite a larger expansion rate.  The anthropic bound on $\Lambda$
scales like $Q^3$, so increasing $Q$ can relax the bound by many
orders of  magnitude \cite{TR, Graesser}.  Similarly, \cite{Graesser2}
showed that allowing the effective Planck scale of a tensor-scalar
theory of gravity to vary also weakens the anthropic bound on $\Lambda$;
this result is potentially important for the string theory landscape.

Despite these shortcomings, one should keep in mind that the
anthropic approach was the only one to predict the range of $\Lambda$
before its nonzero observation, and the value $\Lambda_{\mathrm{obs}}\sim
10^{-123} M_p^2$ is so peculiar from the particle physics perspective
that it seems exceedingly unlikely that a dynamical mechanism could 
by itself explain the observed value of $\Lambda$.  Something like
the anthropic principle therefore appears to be necessary for
understanding the magnitude of the dark energy.  One is thus
motivated to search for some improvement which retains the virtues of
the anthropic principle while getting rid of overly specific
assumptions about the nature of observers.  Such an idea, dubbed the
``causal entropic principle,''  has recently been proposed by
\cite{Bousso}.  It assumes that a good tracer of the potential for
forming structure, and hence observers, is the amount of entropy produced
within a causally connected region (the causal diamond) defined to
start at some early initial time $t_i$, such as reheating, and ending
in the infinite future, which corresponds to a finite
conformal time $\tau$ if $\Lambda >0$.  The created entropy, $\Delta
S$, does not include entropy already present at the moment of
reheating but only that which is created \textit{after} $t_i$.  

The causal entropic principle has several virtues which make it
worthy of further investigation.  First, it predicts a probability
distribution $dP/d\log\Lambda$ such that $\Lambda_{\mathrm{obs}}$ is
within $1\,\sigma$ of the most likely value of $\Lambda$, unlike the
anthropic approach for which $P(\Lambda\le\Lambda_{\mathrm{obs}}) <
10^{-3}$.  Second, it makes minimal assumptions about the detailed
nature of observers or life.  It only assumes that free energy 
(leading to increased entropy) is available, which is a fundamental
thermodynamical requirement for making any measurement.  Third, 
it provides a specific proposal to separate potentially divergent
volume factors (for example, from eternal inflation) into the prior 
probability distribution.  (For recent discussions of volume-weighted 
measures, see \cite{volume}.)
\comment{by
restricting inquiry to a causally connected region, it circumvents
the problem of the divergence of eternally inflating spatial volumes,
which are an obstacle to defining any volume-based measure on the
landscape.   The causal entropic approach thus does not suffer from
ambiguities due to the nonuniqueness of regulating infinite spatial
volumes.}

In their \comment{seminal} paper, the authors of \cite{Bousso}
understandably restricted their attention to the determination of
$\Lambda$.  To carry the idea further, we are interested in exploring
the probability distributions for other cosmological parameters,  for
example $Q$.  This requires knowing how the rate of entropy
production depends on these parameters.  In particular, since 
\cite{Bousso} found that starlight is the main source of entropy,  a
major result of the present work is to show how the rate of star
formation depends on various cosmological parameters.  This was
somewhat roughly estimated in \cite{Bousso}; here we have tried to be
more quantitative.  The most straightforward quantity to vary in fact
is $Q$, and we have studied this dependence to show that the
predicted value of $\Lambda$ in the causal entropic approach is much
less sensitive to $Q$ than in the anthropic approach.  Furthermore,
the entropic prediction for $Q$ is much closer to the observed value
than is the anthropic prediction.  Similarly, we find that the
prediction for the baryon fraction is within a factor of a few of the
observed value.

An important aspect of \cite{Bousso} is the realization that the
typical timescale of star formation and galaxy evolution is of the
same order of magnitude as the age of the universe, providing 
another coincidence problem. The causal entropic view effectively elevates
this coincidence to a principle, with potentially testable 
predictions.\footnote{An interesting paper \cite{Peacock} has recently
analyzed the joint probability of observing a universe with a given $\Lambda$
at a given CMB temperature using star formation and evolution as the key
indicator of observers, following the original ideas of \cite{efstathiou}.
Even though the spirit of \cite{Peacock} is not as general as that of
\cite{Bousso}, the results are, unsurprisingly, similar.}
A process that generates a large amount of entropy on a characteristic
timescale would make more likely, by this principle, a cosmological
constant that becomes dynamically important on a comparable timescale. 
This suggests that there is not a large source of entropy that is acting
on a timescale significantly different than the current age of the universe.
If we discover a new important entropy source, we should expect its 
timescale to be the cosmological one. In fact, such an entropy source would
make the observed value of $\Lambda$ more likely.

Beyond the consideration of universes very similar to our own, the
entropic approach has the capability of quantifying the likelihood of
universes with quite different properties.  This leads one to ask
whether alternative universes with much greater entropy production
than our own can be constructed, addressing the critique of 
\cite{Maor} that existing work has focused too much on universes and life
similar to our own.\footnote{Also, like \cite{Maor},
one might ask whether we should be interested in the conditional 
probability of observing $\Lambda_{\textrm{obs}}$ given the existence of
observers or given the existence of human-like life.  While more traditional
approaches focus on human-like life, the causal entropic principle allows us
to consider more generic observers.}
If so, it might call into
question the validity of entropic principle, since our universe would
then appear to be relatively unlikely.  One way in which entropy might
be copiously generated is through the decay of dark matter. 
Interestingly, hints of excess gamma rays at various energies has led
to speculation that the dark matter in our universe indeed is
unstable.  One goal of this work is to determine whether the entropy
released by such decays can be competitive with that due to starlight;
if so, this could be an indication that the entropic principle can
explain properties of dark matter as well as dark energy.

We will review the causal entropic approach in section \ref{sec2},
including our results for $dP/d\log\Lambda$, and motivate the
improved star formation rate (SFR) which we develop in section
\ref{sec3}.  This will be followed in \ref{sec4} by our results for
the probability distribution for $Q$ and for the baryon fraction. 
Section \ref{sec5} will discuss the production of entropy from
decaying dark matter.  We give conclusions in section \ref{sec6}.
Throughout, except when units are given explicitly, we work with $\hbar=c=1$.
In formulae, we keep factors of $M_P$ and $G$ explicit, but we omit them
for brevity in figures and 
when writing $dP/d\log\Lambda$ (which should be read as 
$dP/d\log (\Lambda/M_P^2)$).

\section{The Causal Entropic Principle}
\label{sec2}
\begin{figure}[htbp]
\includegraphics[scale=0.45]{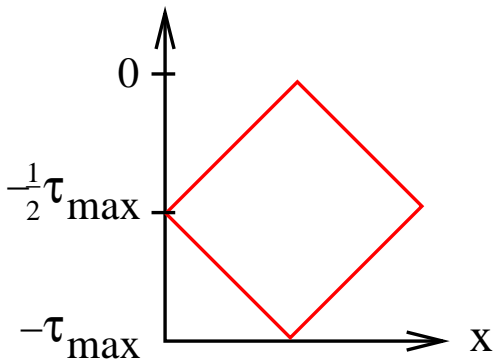}
\caption{The causal diamond}
\label{fig1}
\end{figure}

In this section, we briefly review the basic ideas and results of 
\cite{Bousso}.  The fundamental assumption is that the probability
distribution for $\Lambda$ is proportional to entropy production in
the causal diamond,
\beq
	\frac{dP}{d\Lambda} \propto \Delta S = \int_{t_i}^\infty dt\, V_c(t) 
	\frac{dS}{dV_c dt}
\eeq
where $V_c = (4\pi/3)r(t)^3$ is the comoving volume in the diamond at a given
time, and $dS/dV_c dt$ is the rate of entropy production per
comoving volume.  The causal diamond is defined in conformal time,
$ds^2 = a^2(\tau)(-d\tau^2 + dx^2)$, as the region shown in figure
\ref{fig1}.  The initial time is taken to be the time of reheating,
although results are quite insensitive to this choice since $V_c$ is
negligible at such early times.  In conformal time, the final time
is finite (in fact, it is defined to be $0$)
since $\tau = \int dt/a(t)$ converges for any $\Lambda > 0$;
recall that $a(t) \sim \exp(t\sqrt{\Lambda/3})$ asymptotically.
The initial time is $-\tau_{\mathrm{max}}\approx - 2.8\, t_{\Lambda}^{1/3}
\equiv 2.8\,(3/\Lambda)^{1/6}$.  The diamond achieves a maximum volume
at the intermediate time $-\tau_{\mathrm{max}}$.

Ref.\ \cite{Bousso} estimated the amount of entropy produced by various
astrophysical sources, including active galactic nuclei, supernovae,
and cooling of galaxies,  and determined that starlight,
inelastically scattered by dust, is the largest source of 
entropy.\footnote{The horizon entropy of black holes is a special case which
is excluded; a single black hole's entropy far outnumbers that of
other entropy sources in the universe, but a universe filled with
black holes would seem to be inimical to observers.}
The rate of entropy production per comoving volume can be written
as
\beq
	\frac{dS}{dV_c\, dt}(t) = \int_0^t dt'\, 
	\frac{d^2 S}{dM_\star dt}(t-t')\, \dot\rho_\star(t')\ ,
\eeq
where $(d^2 S/dM_\star dt)(t-t')$ is the rate of entropy production
per stellar mass at time $t$ due to stars born at time $t'$, while
$\dot\rho_\star(t')$ is the rate of stellar mass production (star
formation rate, or SFR) at time $t'$.  $(d^2 S/dM_\star dt)(t-t')$ 
in turn is given by
\beq
\label{d2sdmdt}
	\frac{d^2 S}{dM_\star dt}(t-t') =
	\frac{1}{\langle M\rangle} \int_{0.08 M_\odot}^{M_{\mathrm{max}}
(t-t')}dM\, \xi_{\mathrm{IMF}}(M)\, \frac{d^2 s}{dN_\star dt}\ ,
\eeq
in terms of the initial mass function $\xi_{\mathrm{IMF}}(M)$ (equal to
$0.105\, M^{-2.35}$ for $M\ge 0.5 M_\odot$ and $0.189\, M^{-1.5}$ for
$M<0.5 M_\odot$).  The entropy production rate for a single star,
$d^2 s/dN_\star dt$, 
is its luminosity over an effective temperature, $L_\star/T_{\mathrm{eff}}
\approx (M/M_\odot)^{3.5} \times 10^{54} \textnormal{ yr}^{-1}$.  The
effective
temperature $T_{\mathrm{eff}}\approx 20$ meV reflects the fact that half of
all starlight, originally UV, is scattered into the IR by dust and its
associated entropy is thereby multiplied.  The integration limits
in (\ref{d2sdmdt}) are the minimum and maximum masses of stars of a 
given age, where $M_{\mathrm{max}}(t-t')= \textnormal{max}
(100\,M_\odot,\ [10\textnormal{ Gyr}/(t-t')]^{2/5})$, 
reflecting the fact that the largest stars burn
out more quickly than the smallest ones.

\begin{figure}[htbp]
\includegraphics[scale=0.4]{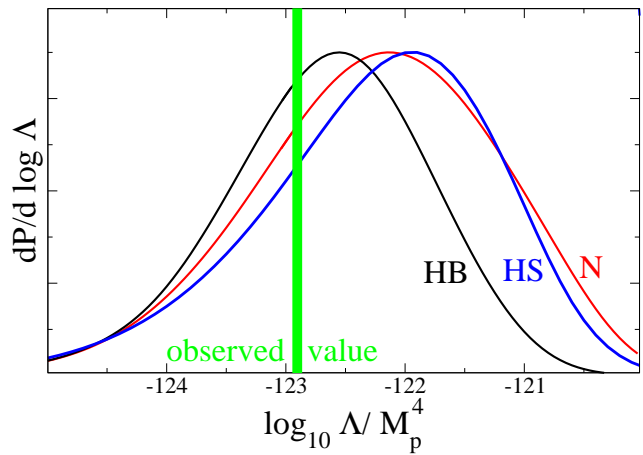}
\caption{Unnormalized probability 
distributions for $\Lambda$ using different star
formation rates \cite{Nagamine,HB,HS}. Vertical line indicates the
observed value of $\Lambda$.}
\label{fig2}
\end{figure}

The total entropy production is thus the convolution of a
cosmological factor, $V_c(t)$, with an astrophysical one, $(dS/
dV_c\, dt)(t)$. The former gives a preference to small values of
$\Lambda$ since $V_c(t)\propto 1/\sqrt{\Lambda}$ and $\int_{t_i}^\infty
dt\,V_c(t)\propto 1/\Lambda$.  The latter function is peaked at a time
$2-3$ Gyr, depending on the choice of the SFR, $\dot\rho_\star$, of which
there are several somewhat different determinations in the literature
\cite{Nagamine,HB,HS}.  The fact that star formation tapers off
in the late universe softens the preference for small $\Lambda$ such
that the total entropy production (and hence by assumption the
probability distribution for $\Lambda$) approaches a constant as
$\Lambda\to 0$, but falls off rapidly as $\Lambda$ becomes much
greater than the observed value.  To better visualize the range of
likely $\Lambda$ values, it is useful to consider the distribution in
$\log\Lambda$, $dP/d\log\Lambda = \Lambda dP/d\Lambda$, which is
peaked at a nonvanishing value of $\Lambda$.  We have reproduced (and
extended) the calculations of \cite{Bousso} to obtain 
$dP/d\log\Lambda$, using three different determinations for the SFR,
Nagamine {\it et al.} \cite{Nagamine} (N), Hopkins-Beacom \cite{HB}
(HB), and Hernquist-Springel (HS) \cite{HS}.  The distributions 
(not normalized) are
shown in fig.\ \ref{fig2}.

Although there is some uncertainty in the prediction for 
$dP/d\log\Lambda$ due to differences in the estimates for the SFR,
all three curves in fig.\  \ref{fig2} show that the observed value of 
$\Lambda$ is within
$1\, \sigma$ of the most likely value, which is a much better agreement
than that obtained using the anthropic approach.  In 
\cite{Bousso}, only the the HB and N SFRs were considered, which are
phenomenologically derived for our universe and in particular for
the observed value of $\Lambda$.  In order to estimate the effect of
varying $\Lambda$, \cite{Bousso} multiplied the entire SFR
by a correction factor $F(t,M)$, taken to be the Press-Schechter
fraction \cite{PS} of matter collapsed into halos at a typical
galactic mass scale $10^7 M_\odot$ and the time $t\sim 2-3$ Gyr when
the SFR is maximized.  (We discuss the Press-Schechter formalism
in some detail in the appendix.)   This is a reasonable, though
somewhat crude approximation, since the true answer must depend on
the details of gravitational collapse over a range of mass and time
scales.  A major goal of the present work is to improve on this
approximation by developing an SFR which has the correct quantitative
dependences on $\Lambda$ and other cosmological parameters.  We have
carried this out, based on an analytical model due to Hernquist and
Springel \cite{HS} (henceforth HS), 
which is the subject of the next section.  Our
result for $dP/d\log\Lambda$ based on the HS SFR, as shown in fig.\ 
\ref{fig2}, is in reasonable agreement with the earlier results. 
More importantly, it allows us to consider with greater confidence
the probability distributions for other parameters, as a way of
further testing the causal entropic principle.

\section{The Hernquist-Springel SFR}
\label{sec3}

As mentioned above, the N \cite{Nagamine} and HB \cite{HB} SFRs are
simple phenomenological formulae derived from observations in our
universe, so they cannot tell us how the variation of cosmological
parameters will affect the SFR.  Fortunately, Hernquist \& Springel
\cite{HS} have developed a simple analytic model for
the SFR based on more detailed  numerical simulations
\cite{astro-ph/0206393}.  Because the HS model can
be written explicitly in terms of cosmological parameters and other
physical constants, it is possible to vary the SFR in reponse to
changes in, for example, the cosmological constant or the amplitude
of cosmological perturbations.  In this section, we will 
briefly describe our implementation of the HS SFR model,
leaving a more detailed discussion of the model for the 
appendix.

HS start with the formula 
\eq{sfr1}
{\dot\rho_\star(t) = \int \frac{dF}{d\ln M} (M,t)\, s(M,t)\,d\ln M\ .}
Here, 
$F(M,t)$ is the 
Press-Shechter fraction (the fraction of matter collapsed into clouds of
mass $M$ or less) \cite{PS},\footnote{HS advocate the use of a
more precise collapse fraction derived by Sheth \& Torman 
\cite{astro-ph/9901122}, but the Press-Schechter fraction
is accurate enough for our purposes.}
and $s=\langle\dot \rho_\star\rangle$ is the averaged rate of star formation
in collapsed haloes. 

The collapsed fraction for a given scale as a function of time
can be simply calculated
from the statistics of Gaussian random fields for an assumed input
matter power spectrum with a power spectrum amplitude that evolves with
time as expected from linear theory. 

The scale at which the collapsed fraction should be calculated is set
by atomic physics: one can collisionally excite hydrogen atomic transitions 
when kinetic temperatures exceed $\sim 10^4$ K. 
The virial relation between mass and 
temperature at time $t$ is found to be 
\beq
M = \frac{(2kT/\mu_m)^{3/2}}{\rho_v}\approx 
\frac{(2kT/\mu_m)^{3/2}}{10 G H(t)}\ ,\label{MTvir}
\eeq
where $k$ is Boltzmann's constant, $\mu_m$ is the mean mass per particle,  
and $M$ depends on the Hubble parameter at the given time. 

Using constants $\delta_c = 1.6868$ and $a = 0.707$, numbers found to 
provide good agreement beween theory and large N-body simulations for
the statistics of collapse,
a simple version of the SFR suggested by HS is 
\beq
\label{SFR}
	\dot\rho_\star = q(t)\left( 1- \mathrm{erf}\left(\sqrt{\frac{a}{2}}\,
\frac{\delta_c}{\sigma_4}\right) 
\right)\ ,
\eeq
where $\sigma_4$ stands for fluctuations at the mass scale which
virializes at temperature $10^4$ K, and the physics of cooling is
parameterized in the prefactor $q(t)$.  

The star formation rate within a collapse object is regulated by the rate of
radiative cooling and the efficiency of radiative cooling is difficult to 
calculate; HS report that a good approximation to the star formation 
efficiency can be characterized by 
\beq
q(t) \propto [ \chi(t) \bar\chi / (\chi(t)^m + \bar\chi^m)^{1/m} ]^{9/2\eta}\ ,
\label{qt}
\eeq
where $\chi=(H/H_0)^{2/3}$, and $\bar\chi=4.6$, $\eta=1.65$, and $m=6$
provide good fits to numerical simulations ($H_0$ is the observed value
of the Hubble rate taken as a reference value).  Because we are interested
in relative probabilities, we do not need to normalize the star formation 
rate.

\begin{figure}[htbp]
\includegraphics[scale=0.4]{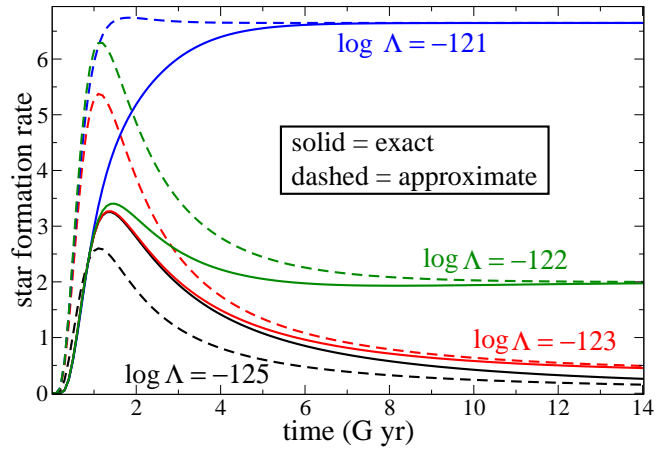}
\caption{Solid lines: the star formation rate used in this paper, for
several values of $\rho_\Lambda$.  Dashed lines are the approximation 
(\ref{hsapp}).  Common proportionality constants have been omitted.}
\label{fig3.5}
\end{figure}

To illustrate the parametric dependences of our modified SFR, we 
plot $\dot\rho_\star(t)$ for several values of $\Lambda$ in figure
\ref{fig3.5}.  We also show there a naive approximation to the 
HS result, which consists of the analytic fit which HS made to
their result,
\beq
\dot{\rho}_{\star,\mathrm{approx}}(t) \propto
\frac{\chi^2}{(1+\alpha(\chi-1)^3\exp(\beta\chi^{7/4})}\ .
\label{hsapp}
\eeq
Here all dependence on the cosmological parameters is due to 
$\chi$, which depends upon $\Lambda$ through the time-dependent Hubble
rate, and the
parameter $\beta$, which depends on the density contrast $Q$ as 
$\beta\sim Q^{-2}$ via the Press-Schechter formalism.  We have
normalized (\ref{hsapp}) to agree with the more exact result at late
times.  The rough approximation tends to overestimate or underestimate
the peak value of the more accurate SFR.  

\begin{figure}[htbp]
\includegraphics[scale=0.4]{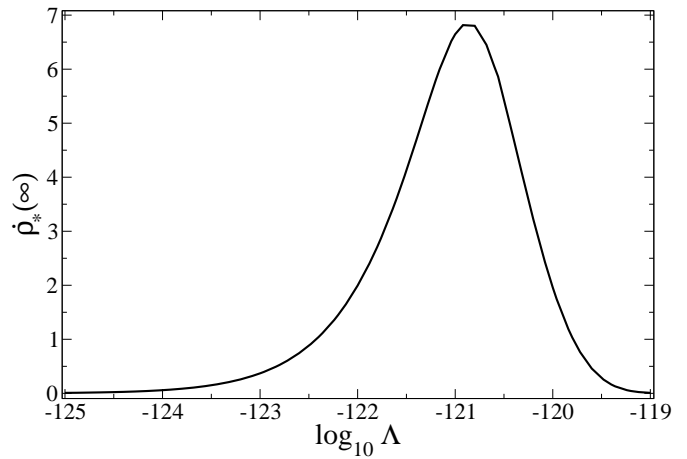}
\caption{Asymptotic value of the star formation rate at large
times, as a function of $\Lambda/M_P^2$.}
\label{fig3.6}
\end{figure}

Let us also contrast our treatment with that of \cite{Bousso},
where the shape of the SFR was assumed to remain constant, and only
its overall normalization depended on $\Lambda$.  Our results
indicate a significant change in shape with $\Lambda$, with the peak
even disappearing at large $\Lambda$.  The reader may be surprised
that the SFR increases with $\Lambda$, since increasing the expansion
rate is supposed to delay formation of structure.  However since the
effect of $\Lambda$ appears in the prefactor of $\dot\rho_\star$, through
$\chi$ (see eqs.~(\ref{qt}) or (\ref{hsapp})), there is initially an
increase with $\Lambda$.  The decrease predicted by Press-Schechter
becomes apparent for $\log(\Lambda/M_P^2) \gsim -121$.  In
figure \ref{fig3.6} we show this dependence by plotting the large-time
asymptotic value of $\dot\rho_\star$ as a function of $\Lambda$.  This
function is peaked around $\log(\Lambda/M_P^2) \gsim -121$.  Taken by itself
it would indicate a lower probability for our universe with
 $\log(\Lambda/M_P^2) \sim -123$ than the full entropic treatment gives.

In the above we have given the simplest reasonable implementation of
the HS model, and there is room for improvement, as
outlined in the appendix.  However, this is a physically motivated
model for how star formation depends on cosmological parameters, and
it captures the important elements that any such model must have.

\section{Predictions for other parameters}
\label{sec4}

As long as the properties of the universe are such that starlight
continues to be the dominant source of entropy production, we now
have the tools to compute $\Delta S$ as a function of cosmological
parameters beyond just $\Lambda$.  These include the density contrast
$Q$, the ratio of baryonic matter to dark matter, the ratio of matter
to photons, and possibly the spatial curvature.  The latter is more
subtle to try to quantify because it is time-dependent.  One must
specify the curvature at some reference time, for example at
matter-radiation equality.  However, there are good reasons for
believing that the curvature of our universe was determined by
inflation rather than by environmental considerations.  We therefore
will confine our investigations to the explicitly time-independent
quantities.  

\subsection{Density contrast}
The most straightforward cosmological parameter to vary is the density
contrast $Q$, since it appears only through (\ref{sigmaeq}).
It is also interesting because of the great sensitivity of the
anthropic prediction for $\Lambda$ on $Q$.  This can be understood
from (\ref{Geq}) and (\ref{sigmaeq}), which show that $Q$ and
$\Lambda$ appear in the Press-Schechter fraction through the
combination $Q \Lambda^{1/3}$.  In the anthropic approach, structure
formation is the most important effect, and this causes the anthropic
upper bound on $\Lambda$ to scale like $Q^3$.  In contrast, the
entropic approach gives less weight to very early structure formation
due to larger $Q$, because the volume of the causal diamond is smaller
at earlier times.  One can thus anticipate less sensitivity to $Q$ in
the entropic approach.  

\begin{figure}[htbp]
\includegraphics[scale=0.4]{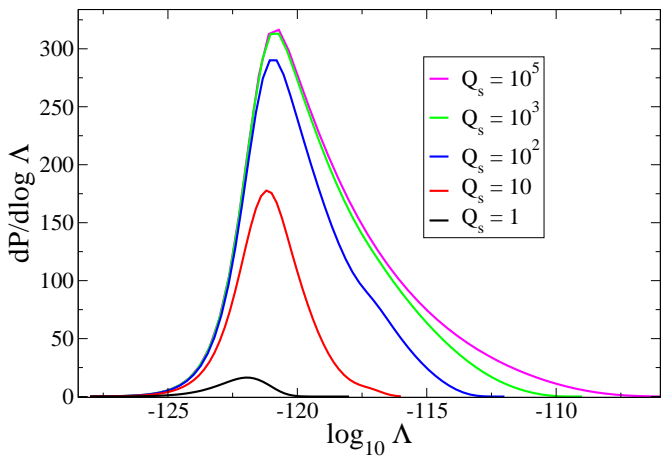}
\caption{Unnormalized 
probability distributions for $\Lambda$ at different values
of $Q = Q_s Q_{\mathrm{obs}}$}
\label{fig3}
\end{figure}

This expectation is borne out in figure \ref{fig3}, where $Q$ is
varied by 4 orders of magnitude above the observed value,  $Q = Q_s
Q_{\mathrm{obs}}$.  In fact, the curves for $dP/d\log\Lambda$
saturate near the highest one shown as $Q\to\infty$.  The  position
of the peak of the distribution shifts by only an order of magnitude,
which is a much weaker dependence than the $\Lambda\sim Q^3$
dependence due to structure formation alone.  As $Q$ is varied, the
epoch of star formation moves to earlier times, but the
characteristic time of 1 Gyr remains relevant. It matters little
whether stars form at $10^6$ years or $10^8$ years since their
characteristic lifetime is $10^9$ years. At extremely low $Q$,
though, the $\Lambda \sim Q^3$ dependence should reemerge.

\begin{figure}[t]
\includegraphics[scale=0.4]{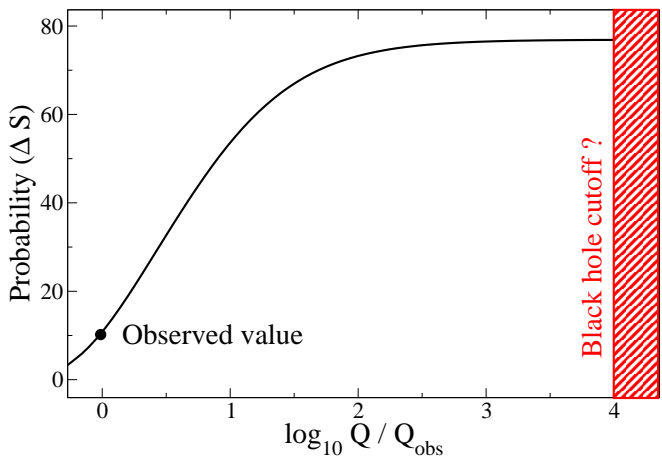}
\caption{Unnormalized probability distribution for $Q$ at fixed
$\Lambda=\Lambda_{\mathrm{obs}}$.}
\label{fig4}
\end{figure}

\begin{figure}[t]
\includegraphics[scale=0.53]{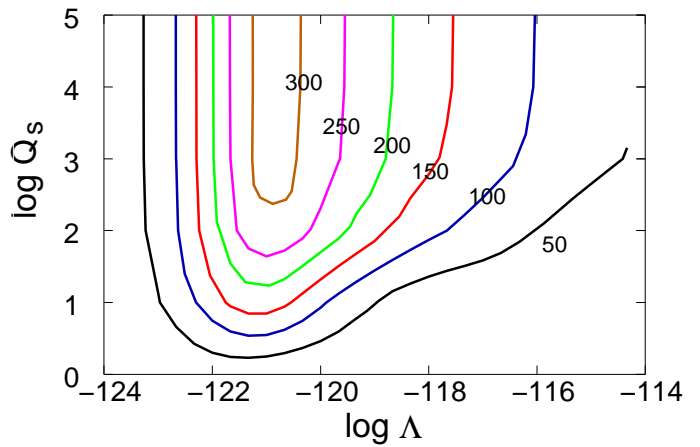}
\caption{Contours of the probability distribution $dP/d\log\Lambda$,
varying $Q$ and $\Lambda$.}
\label{contour}
\end{figure}

We also plot $\Delta S(Q)$ for fixed $\Lambda =
\Lambda_{\mathrm{obs}}$ in figure \ref{fig4} and a contour plot
varying both $\Lambda$ and $Q$ in figure \ref{contour}.  Although our
universe is not at the peak of the distribution, the probability of
$Q_{\mathrm{obs}}$ is only  a factor of $8$ smaller than the most
likely value.  For comparison,  the usual $\chi^2$ statistic is of
order $-2 \ln P$, so a factor of 8 corresponds to a change in the
effective $\chi^2$ on the order of 4.2. 

One effect which has not been quantified in our approach is the
cutoff on $Q$ which would result from overproduction of black holes;
beyond some large value of $Q$, formation of stars (and observers)
would be impeded by the loss of material to black holes.  A
discussion of this issue can be found in \cite{Tegmark}.

The reader should note that we have only calculated the entropic weighting
factor for the probability, independent of any prior distribution for $Q$.
What our results show is that, first, the observed value of $\Lambda$ is 
still near the most likely value, 
even for large values of $Q$, and also that entropic
effects will not modify the prior distribution for $Q$ by more than an order
of magnitude across a large range.  It is reasonable to expect the prior
distribution for $Q$ to be $dP/dQ \propto Q^n$ with $-1\lsim n\lsim 1$
\cite{Graesser}; the smaller value works well with our results.

\subsection{Baryon fraction}

Another cosmological parameter which can obviously affect star
formation is the ratio of baryonic to dark matter, or baryon fraction
$f_b$.  Since stars are baryonic, the SFR should have an overall
factor of $f_b$ to account for the availability of baryonic material
for star formation, as appears in (\ref{cooling2}).  

The details of structure formation are also
affected by $f_b$; at very large values, gravitational collapse is
impeded by pressure; this is known as Silk damping. 
Because the baryonic matter cannot collapse easily due to pressure,
the matter power spectrum in (\ref{linear}) scales as $(1-f_b)$
for a fixed initial amplitude $Q$, so the r.m.s.~fluctuation $\sigma$ scales
like $(1-f_b)^{1/2}$.  

\begin{figure}[h]
\includegraphics[scale=0.4]{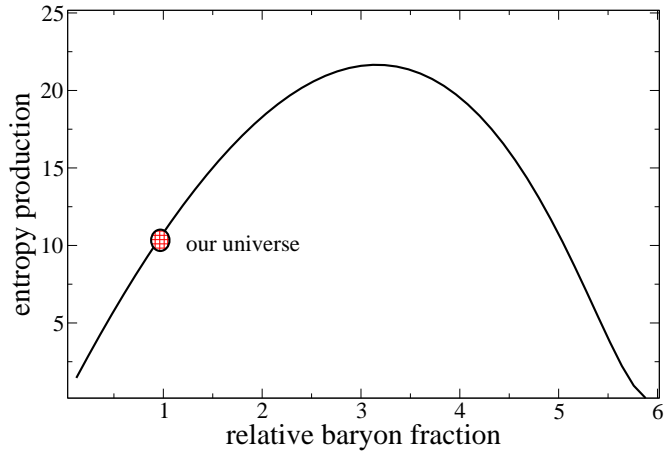}
\caption{Unnormalized probability distribution for relative baryon fraction 
$r_b = 6f_b$ at fixed $\Lambda=\Lambda_{\mathrm{obs}}$.}
\label{fig5}
\end{figure}

Since our universe has $f_b\sim 1/6,$ we define the relative baryon 
fraction as $r_b = 6f_b$.  The dependence of $\dot\rho_\star$ on $r_b$ as
described above can be summarized as making the replacement
\beq
\dot\rho_\star(\mu,\,\sigma) \to r_b\,\dot\rho_\star\left(\mu,
\,\left(\frac{6-r_b}{5}\right)^{1/2}\sigma\right)\ .
\eeq
 The entropic distribution for $r_b$
is shown in figure \ref{fig5}.  Similarly to the distributions for
$\Lambda$ and $Q$, it indicates that the measured value of $f_b$ is
not quite the optimal one, but also not very far from being so: the
probability of $f_{b,\mathrm{obs}}$ is $\sim 1/2$ of that for the most likely
value.

The reader should note that, as with $Q$, we have not attempted to
calculate the prior probability distribution for the baryon
fraction.  Our results show that the entropic weighting does not
change the probability distribution by more than a factor of 2 for
$1/6\lsim f_b\lsim 5/6$.  Therefore, the value taken in our universe
is reasonably probable unless the prior distribution is strongly 
peaked at some other value.

\subsection{Matter abundance}
\begin{figure}[h]
\includegraphics[scale=0.4]{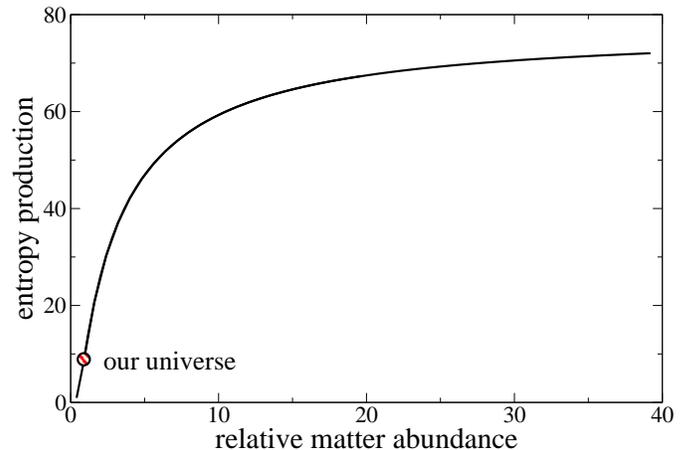}
\caption{Unnormalized probability distribution for relative matter abundance
 $r_m$ at fixed $\Lambda=\Lambda_{\mathrm{obs}}$.}
\label{fig6}
\end{figure}

A third quantity which can strongly affect star formation is the
overall abundance $\xi$ of matter relative to photons, for fixed baryon
fraction, $Q$ and $\Lambda$.  In a universe with larger $\xi$, matter
domination will occur earlier, and the CMB temperature at a given
time will be reduced.  In our analysis, radiation does not play a
siginificant role in the evolution of the scale factor of the
universe, so we can again focus on how $\xi$ affects the SFR.
The dependence of the SFR on $\xi$ has
two origins; first, $\mu$ scales as $\xi^2$ by definition.  Second,
the linear growth factor for perturbations,  
$G(a_{eq})$ in (\ref{growthD2}), depends on $\xi$ simply because it is
evaluated at matter-radiation equality.  As long as matter dominates over
curvature and $\Lambda$ at matter-radiation equality, it is 
straightforward to find \cite{Tegmark} that $G(a_{eq})\sim \xi^{-4/3}$.
From those arguments, we can infer that the $\xi$
dependence comes in the form
\beq
\dot\rho_\star(\mu,\,\sigma) \to \dot\rho_\star\left(r_m^2\mu,
\,r_m^{4/3}\sigma\right)\ ,
\eeq
where the relative matter abundance $r_m$ is the rescaling of $\xi$
such that $r_m=1$ for our universe.  Unlike the case of varying
$f_b$, there is no reduction in the matter power spectrum as $r_m$ is
increased, so star formation and entropy production only become
more efficient as $\xi$ increases but in a manner qualitatively
similar to the dependence on the density contrast $Q$.  
This is shown in figure \ref{fig6}.  Similarly to the case of $Q$, the
relative probability of the observed value of $\xi$ is $\sim 1/8$.

Again, we have not attempted to calculate the prior probability distribution
for $\xi$, but our results show that the entropic weighting factor itself
does not disfavor our universe greatly.

\section{Decaying dark matter}
\label{sec5}

In the previous section we confined our attention to the variation of
standard cosmological parameters.  More generally, one could imagine
universes with properties rather different from our own, requiring
for their description other parameters than those characterizing the
standard cosmological model.  A possible new source of entropy is the
decay of dark matter particles.  We shall show in this section that
one can design universes not so different from our own where entropy
production is actually dominated by dark matter decays.

Interestingly, there are experimental hints from observations of
gamma rays that dark matter could be unstable.  The EGRET collaboration
observes $\gamma$-rays in the $2-10$ GeV range which are unaccounted
for by standard mechanisms \cite{Strong}.  It has been suggested that
these excess $\gamma${}s are due to the annihilation \cite{deBoer} or
decays \cite{Buchmuller} of dark matter.  At lower energies, the 
COMPTEL experiment observed an excess in the $1-5$ MeV range.  Ref.\
\cite{Cembranos} has proposed the decays of Kaluza-Klein (KK) or
supersymmetric (SUSY) dark matter to explain this anomaly.  In a
similar energy range, numerous experiments, including the SPI
spectrometer on the INTEGRAL observatory \cite{Jean}  have observed 511 keV
photons from the galactic center, indicative of an excess of positrons
annihilating nearly at rest.  Astrophysical explanations have not
convincingly accounted for this excess, leading to suggestions that
the positrons are a result of dark matter annihilations \cite{Boehm}
or decays \cite{decays}.

\begin{figure}[h]
\includegraphics[scale=0.39]{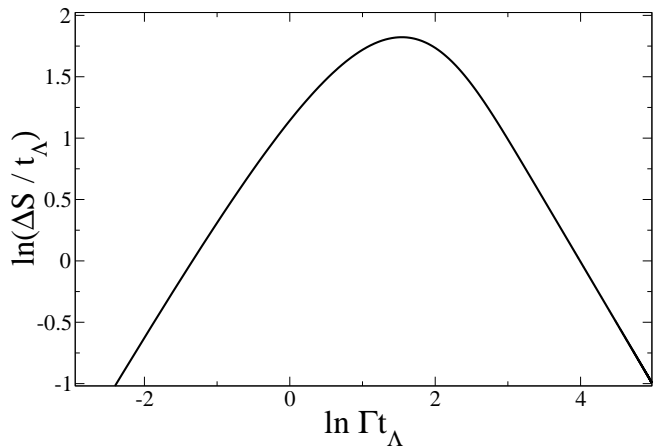}
\caption{Entropy produced by dark matter decays as a function of the
dimensionless combination $\Gamma t_\Lambda$ (upto an overall 
proportionality).}
\label{fig7}
\end{figure}

If indeed dark matter is shown to be unstable, it is curious that the 
MeV-scale anomaly requires a lifetime which is within a few orders of
magnitude of the age of the universe.  This could then pose a new
coincidence problem.  Interestingly, the entropic principle could
explain such a coincidence, as we now discuss.  The rate of entropy
production from dark matter (DM) decays is much simpler to calculate than
that from stars; it is given by
\beq
\frac{dS_d}{dV_c dt}= g_s\,\Gamma\, n\, e^{-\Gamma t}\ ,
\eeq
where $\Gamma$ is the DM decay rate, $n$ the DM density, and $g_s$ the
entropy increase per particle decay.  We find analytically that the
entropy produced by decays has the functional form
\beq
\Delta S_d = t_\Lambda\, f(\Gamma t_\Lambda)\ ,
\eeq
where we recall that $t_\Lambda \equiv \sqrt{3/\Lambda}$.  The
function $f$, calculated numerically, 
is shown in figure \ref{fig7}.  It is closely fit by the analytic
approximation
\beq
\ln f \approx c_1 + c_2 x - \ln(1+c_3 e^{c_4 x})\ ,
\eeq
where $x=\ln\Gamma t_\Lambda$ and the constants are $c_1=1.164$,
$c_2 = 0.0512$, $c_3 = 0.889$, and $c_4=1.921$.  
The distribution is
peaked for a DM lifetime given by
\beq
\tau = \frac{1}{\Gamma} \approx \frac{t_\Lambda}{4.7}\ .
\eeq
In our universe, $t_\Lambda = 16.7$ Gyr, corresponding to a preferred
lifetime of 3.6 Gyr.   This could then explain why dark matter should
be decaying on a timescale comparable to the age of the universe.

The above argument might only be valid if the entropy produced by
decays is greater than or equal to that coming from starlight.  We
must therefore compare the magnitudes of $\Delta S$ from decays and
from stars.  Ref.\ \cite{Bousso} found that the peak rate of entropy
production by stars was of the order
\beq
\frac{dS_\star}{dV_c dt} = 10^{63}\, \textnormal{Mpc}^{-3}\, 
\textnormal{y}^{-1}\ .
\eeq
Using the known energy density of dark matter, we find that the
corresponding expression from decays is
\beq
\frac{dS_d}{dV_c dt} \approx 10^{66}g_s\! \left(
\frac{\Gamma}{\textnormal{Gyr}^{-1}}
\right)\!\! \left(\frac{\textnormal{eV}}{m}\right)\!\!
\left(\frac{\rho_{\mathrm{DM}}}{(10^{-3}\mathrm{eV})^4}\right) \!
\mathrm{Mpc}^{-3}\, \mathrm{y}^{-1}\ ,
\eeq
where $m$ is the DM mass.  Therefore the requirement for entropy from
DM to dominate is
\beq
\frac{m}{g_s} \le \textnormal{\ keV}\ .
\eeq
This can obviously be satisfied even if $g_s\sim 1$ if dark matter is
sufficiently light.\footnote{Notice that $g_s$ is quantized, so it
cannot be arbitrarily small and still nonzero.}  However, to explain the 511
keV signal, for example, one needs $m\sim$ MeV, which would require
$g_s \sim 1000$. At first sight, this would seem like an unreasonably
large increase in entropy per decay.  However, it might actually be
easy to achieve, when one considers that electrons are typically
produced along with the positrons, since we know that dark matter is
charge-neutral, and that these electrons, if even mildly relativistic,
must produce a great number of lower-energy photons as they
thermalize.  

Let us consider how an electron of energy $\sim$ MeV thermalizes
within a galaxy.  Synchrotron radiation is kinematically blocked
because the cyclotron frequency is only $ eB/m_e \sim 10^{-13}$ eV
for $B\sim 10^{-9}$ T, while photons have a plasma mass of
$m_\gamma\sim (\alpha n_e / m_e)^{1/2} \sim 10^{-12}$ eV, since
$n_e\sim 1$ cm$^{-3}$.   Instead, the electron loses energy to the
galactic medium by Coulomb interactions; using the formalism
developed by \cite{coulomb}, we can see that an electron would lose all its
kinetic energy over $\sim 10^5$ yr and $\sim 3$ kpc.  
Thermalization thus takes place much faster than
the age of the universe.  The final energy is of the order $E\sim 1-100$ eV,
resulting in an entropy production of $\Delta S\sim$ MeV$/E\sim
10^{4}-10^6$.  For electrons produced in the galactic halo,
thermalization will be less efficient.  Nevertheless, from this number
we see that it is possible for DM decays which produce mildly
relativistic $e^+ e^-$ pairs to result in entropy production which is
similar in magnitude to that produced by stars.

It might be objected that the existence of such low energy photons
should be irrelevant for creating observers, even if they far
outnumber photons originating from stars.  However we wish to avoid
making any assumptions about the detailed nature of observers, since
this was one of the arbitrary features of the anthropic principle
which one would like to overcome with the entropic principle.

\section{Conclusions}
\label{sec6}

Using the analytic star formation model of HS, we have improved upon
and extended the calculations of \cite{Bousso} to test the causal
entropic principle further.  First, we recalculated the probability
distribution  for the cosmological constant  $\Lambda$ alone, finding
good agreement with the approximations of \cite{Bousso}.  We
subsequently extended the original analysis to allow for simultaneous
variation of $\Lambda$ with the density perturbation amplitude $Q$,
finding that the most probable value of $\Lambda$ changes by
less than an order of magnitude as $Q$ varies by 5 orders of
magnitude.   In addition, the width of the probability distribution
grows as $Q$ grows, so the observed value of $\Lambda$ remains
reasonably probable even at large $Q$.  The observed value of $Q$ (at
fixed $\Lambda$) has about $1/8$ the probability of the peak values,
which may or may not be compensated by the prior probability
distribution. Due to the flexibility of the HS SFR, we have also been
able to calculate the entropic weighting factor for varying baryon
fraction and matter abundance, finding that the observed values are
not unreasonable.

We have moreoever demonstrated the adaptability of the causal
entropic principle to universes which could be qualitatively
different from our own, by considering entropy produced by the decay
of massive particles.  If entropy due to particle decay is the
dominant source of entropy production, we find that the particle
lifetime should be about 20\% of $t_\Lambda$ ({\it i.e.} 3.6 Gyr,
since $t_\Lambda=16.7$ Gyr in our universe) for the optimal increase
in entropy.  We showed that in fact it is easy for  entropy from
decays to dominate that from stars if the dark matter is warm
($m\sim$ keV).  Although warm dark matter seems to be ruled out in
our universe, a small warm component could still generate significant
entropy through its decays.  Intriguingly, even decaying MeV dark
matter, hinted at by various observed excesses in the galactic and
diffuse gamma ray spectra, could produce more entropy than stars 
if its lifetime is comparable to the age of the universe.

In summary, we have tested the causal entropic principle by varying a
number of parameters which could be environmental in nature, as
opposed to being fundamental constants, and it has so far survived
these tests.  More ambitiously, one could consider particle physics
parameters like the masses and charges of electrons and protons
varying in a landscape of vacua of the fundamental underlying
theory.   Even further afield, we can imagine universes with
different gauge groups and particle content.  It may be interesting
to consider how the entropy produced in such universes compares to
our own, to get a better idea of whether the entropic principle works
in a broader setting.

\begin{acknowledgments}
We are grateful to G.\ Kribs, R.\ Harnik and
C.\ Wetterich for helpful discussions.  We thank A.\ Linde for
critical comments on the issue of volume-based versus other measures
of the probability density and M.~Salem for discussion about whether the
anthropic principle should be concerned with human-like or generic
observers.  Our research is supported by
the Natural Sciences and Engineering Research Council of Canada.  
G.H.\ also acknowledges the Canadian Institute for Advanced Research.
A.F.\ is additionally supported by the IPP \& PI.
\end{acknowledgments}

\appendix*
\section{Details of the Hernquist \& Springel SFR}

In this appendix, we review in more detail the HS SFR, which was developed
based on more detailed  numerical simulations
\cite{astro-ph/0206393} combined with analytic reasoning.  

We remind the reader of the basic formula 
\eq{sfr1_A}
{\dot\rho_\star(t) = \int \frac{dF}{d\ln M} (M,t)\, s(M,t)\,d\ln M\ .}
Here, 
$F(M,t)$ is the 
Press-Shechter fraction (the fraction of matter collapsed into clouds of
mass $M$ or less) \cite{PS},
and $s=\langle\dot \rho_\star\rangle$ is the averaged rate of star formation
in collapsed haloes.  We can now break down the individual parts of this
formula.

\subsection{Virialization}

An important issue for star formation is the virialization of
collapsing gas, which we here review,  following the discussion of
\cite{Liddle:2000cg} on the spherical collapse model in a
matter-dominated FRW universe.

Consider a universe of critical density in matter with a spherical overdensity
(and a corresponding underdense shell).  There is an exact solution for the
evolution of this universe starting from $a=0$ at proper time $t=0$ (its  
precise form is unimportant for us).  The
overdense sphere reaches its turn-around radius at time $t_{ta}$
and then recollapses completely (to a local value $a=0$) in time
$2t_{ta}$.  At the time of complete 
recollapse, the overdensity in the linearized theory is $\delta_c=1.686$, 
which is known as the critical overdensity. Any fluctuation in the initial
density field with a larger linearly evolved overdensity would have 
collapsed at an earlier time.

Of course, the overdense sphere does not recollapse completely,
due to the effective pressure supplied by the random velocities
that the particles would inevitably acquire during the collapse when
deviations from strict spherical symmetry are allowed.  
From the virial theorem, energy  redistributes
among the dust particles such that the total kinetic energy becomes
$-1/2$ the total potential energy in the final state.  This suggests
that the virial radius $R_v$ be $1/2$ the turn-around radius in the
matter-dominated universe.   Therefore, between turn-around and 
virialization, the density of the sphere increases by $8$. 
Meanwhile, since the time doubles, the background density decreases
by a factor of $4$.  At turn-around, the density contrast is
$\rho/\bar\rho = 9\pi^2/16$, so the virial density becomes $\rho_v =
18\pi^2 \rho_m$.  

Other virial quantities can then be determined in terms of the virial density
and the total mass of the collapsed cloud.  Ignoring order-unity numerical
factors, the virial radius for a mass $M$ is $R_v = (M/\rho_v)^{1/3}$, and
the virial velocity of matter particles is 
$V_v=(GM/R_v)^{1/2}=(G^3M^2\rho_v)^{1/6}$.  Finally, the virial temperature
for a given mass is
\eq{tvir}{T_v = \frac{\mu_m}{2k} V_v^2 \sim \frac{\mu_m G}{2k} 
\left(M^2\rho_v\right)^{1/3}\ ,}
where $\mu_m\approx 0.6 m_p$ is the 
(appropriately averaged) molecular mass in the cloud.
These are all time dependent through the evolution of the background
densities, and the last relationship gives rise to (\ref{MTvir}).

A slight generalization of the above argument allows the calculation
of the virial density in other backgrounds.  For instance, the same
collapsing  solution can be used to determine the virial density when
the background  universe has curvature; however, the virial density
is determined by  comparison to the spatially curved background.  An
analytic approximation to the exact virial density is given in 
\cite{astro-ph/9710107}, based on  a derivation in 
\cite{Lacey:1993iv}.  In the case of a background universe with a
cosmological constant (but no spatial curvature), the evolution of
the overdense sphere must be calculated numerically, but the virial
density is derived in much the same way.  One caveat is that the
cosmological constant modifies the gravitational potential, so the
virial radius is no longer half of the maximum radius of the
overdense cloud.  Based on calculations in 
\cite{astro-ph/9601088}, \cite{astro-ph/9710107} found an
approximate analytical formula for the virial density in terms of
the background densities. We have calculated the virial density in
terms of the background densities for a universe with three
components, matter, curvature, and $\Lambda$.

In all the cases, however, the virial density is within two orders of
magnitude of the dominant component of the background density.  Since
the virial  density only enters our calculations through a fractional
power, the precise value is not terribly important, and we simply
use the result for the matter-dominated universe. (We plan to correct
this detail in future work.)

\subsection{Press-Schechter Collapsed Fraction}

The Press-Schechter collapsed fraction \cite{PS} is the fraction of matter
bound in halos of mass $M$ or less, which one can approximate as the
fraction of density fluctuations at length scale $R$ with density contrast
greater than the critical density $\delta_c$.  It is given by
\eq{ps}{F(M,t)=\textnormal{erf}
\left(\frac{\delta_c}{\sqrt{2}\sigma(M,t)}\right)\ ,} 
where $\sigma$
is the root-mean-square density fluctuation with a wavelength  given
by the comoving radius associated with the mass scale $M$: 
$R^3=(3/4\pi)(M/a^3\rho_m)$.

Therefore another quantity needed for determining the SFR is the linear
growth of fluctuations in the argument of the Press-Schechter function.  
In linear theory, the variance in overdensity in spheres that enclose 
a mass $M$ is
\eq{linear}{\sigma^2(M,a) = D(a)^2\int_0^\infty \frac{dk}{2\pi^2} k^2
P(k) \left[\frac{3 j_1(kR)}{kR}\right]\ .} 
This form factorizes the
variance into an overall linearized growth factor and a matter power
spectrum, which is valid after the end of radiation domination.  It
does not work during radiation domination because modes inside and
outside the horizon evolve differently during that time.  After the
radiation era, $P$ takes into account the different evolutions. There
are additional small effects after radiation domination but before
decoupling that we will neglect here. 

One can then integrate 
\eq{spectrum}{\int_0^\infty \frac{dk}{2\pi^2}
k^2 P(k) \left[\frac{3 j_1(kR)}{kR}\right] \equiv Q^2 \, \Sigma(\mu)^2 \
,} 
where the dimensionless parameter $\mu=\xi^2 M/M_P^3$ is
proportional to the mass in terms of the horizon mass at
matter-radiation equality \cite{Tegmark}. Here 
$\xi=\rho_m/n_\gamma$ is a time-independent measure (apart from decays
which increase the density of photons)
of the amount of nonrelativistic matter in the universe. 
Empirically,  the spectrum is found to be well approximated by 
\cite{Tegmark} 
\beqa 
\Sigma(\mu) &=&
\left[\left(9.1\mu^{-2/3}\right)^{-0.27}\right. \nonumber\\
&+&\left.\left(50.5\log\left(834+
\mu^{-1/3}\right)-92\right)^{-0.27}\right]^{-1/0.27} 
\label{spectrum2} 
\eeqa 
and $Q = \delta\rho/\rho$ is the amplitude of fluctuations at
horizon entry, which coincides with the value at the end of inflation.

Next, we compute the linear growth factor, assuming that matter 
dominates over both curvature and $\Lambda$ during the radiation
era.  If there were a single overall growth factor
during the radiation era, we would have for  sub-horizon
scales at sufficiently early times that $a<a_{eq}$
\eq{growthD}{D(a) = 1+\frac{3}{2}\frac{a}{a_{eq}}\ .} This is a
well-known result (see for example \cite{Liddle:2000cg}).  After the 
radiation dominated era ($a>a_{eq}$), (\ref{growthD}) generalizes to
 \cite{Tegmark}
\eq{growthD2}{D(a) =
1+\frac{3}{2}\frac{G(a)}{G(a_{eq})}\ ,} 
where the
spectrum $\Sigma(\mu)$ accounts for the differential growth of
different wavelengths during the radiation era. Shortly after
the end of the radiation era, the first term becomes unimportant. The
well-known result for matter, curvature, and/or $\Lambda$-dominated 
universes is
\eq{growthG}{G(t) = H(t) \int_0^{a(t)} \frac{da}{(aH)^3}} as long as the
universe is not pure de Sitter 
\cite{Liddle:2000cg} or dominated by a form of dark energy that is not
a simple cosmological constant.   
The authors of \cite{HS} obtain an analytic approximation for
(\ref{growthG}) for their SFR, but that approximation is not 
valid for values of 
$\Lambda$ much different that in our universe.  
Therefore we evaluate the integral numerically.  

Nevertheless, it is enlightening to have an analytic
formula to get some intuition for parametric dependences.  Ref.\
\cite{Tegmark} finds that the growth factor for
a flat universe is approximated by
\beq
	D(t) \approx 0.2 \frac{\xi^{4/3}}{\rho_{\Lambda}^{1/3}}
	x^{1/3} \left[ 1 + \left(\frac{x}{1.44}
\right)^\alpha\right]^{-1/3\alpha}\ ,
\label{Geq}
\eeq	
where $\alpha = 0.795$, 
$\rho_\Lambda$ is the vacuum energy density, and 
$x = \rho_\Lambda/\rho_m(t)$ serves as a dimensionless ``time'' variable.

\subsection{Cooling and Star Formation Efficiency}

The other key factor in the HS analysis is the star formation rate in 
a collapsed halo of mass $M$, denoted by $s(M,t)$. 
HS propose a simple model for gas cooling, 
which matches well with numerical studies of a detailed multi-phase model 
\cite{astro-ph/0206393}.  They find that
 the star formation rate is  proportional to
the radiative  cooling rate\comment{, though with a time-independent 
``constant'' of proportionality}.
This model ignores metal-line cooling and the slow change of elemental
abundances due to star burning.  For our purposes, metal-line cooling
is  sufficiently small to neglect, while the change of abundances
(nuclear  fuel depletion) is important only for special values of
the cosmological parameters, so we will ignore both of these effects
in this paper. This also neglects cooling through $H_2$ molecules, a
mechanism which may be important in forming the first stars in our
universe but is believed to be currently negligible.

From their numerical studies, HS argue
for a factorization
\eq{sf1}{s(M,t) = q(t)\times \left\{\begin{array}{cc} 0, & T_v<10^4 
\mathrm{\, K}\\
1/3, & 10^4 \mathrm{\, K}< T_v<10^{6.5} \mathrm{\, K}\\ 1, 
& 10^{6.5}\mathrm{\, K}<T_v\end{array}\right.}
and calculate the cooling rate at $10^7$ K.  Based on numerical 
results and physical reasoning about winds in star forming regions, HS 
assume a simple form for the proportionality factor,
\eq{cooling1}{q(t)\propto \min\left( \frac{1}{t^\star}, \bar q(t)\right)\ .}
The idea is that at high densities, the star formation rate
saturates at (slightly less than or equal to) 
the gas consumption time scale $t^\star$, approximately $2.1$ Gyr.
Then $\bar q$ is the star formation rate due to radiative processes, calculated
without regard to this upper limit.

To compute the cooling rate, consider the cooling time at a radius $r$ from the
center of the cloud.  This is
\eq{tcool}{t(r) = \left(\frac{3}{2}kT\right) \left(\frac{\rho_b(r)}{\mu}\right)
\left(\frac{1}{n_H^2\Omega(T)}\right)\ ,}
where $\rho_b$ is the baryon density within the cloud, $n_H$ is the hydrogen
number density, and $\Omega(T)$ is the cooling function.
We can understand this formula as follows: $3kT/2$ is the energy per 
particle, $\rho_b/\mu$ is the particle number per volume, $n_H^2$ is 
proportional to the rate of hydrogen collisions, and $\Omega(T)$ is 
proportional to the energy loss per collision.  To phrase this in
terms of more
fundamental parameters, we write $n_H=X\rho_b/m_h$, where $X$ is the hydrogen
mass fraction.  Following HS, we take the density of the cloud to be
a power law
\eq{clouddensity}{\rho_b(r) = \frac{(3-\eta)f_b M}{4\pi R_v^{3-\eta} r^\eta}
\ ,}
where $f_b=M_b/M$ is the baryon mass fraction.  HS note that 
$\eta=2$ is the prediction for a 
thermal distribution, but matching to  numerical results gives
$1.5<\eta<2$ with $\eta=1.65$ as the best fit value.

From the above, we can show that
\eq{drdt}{\left.\frac{dr}{dt}\right|_{\mathrm{cool}} = \frac{1}{dt/dr} = 
\frac{1}{\eta}\frac{r}{t}\ .}
Moreover, HS argue that the cooling time should be the natural 
timescale for the gas to 
virialize, $t=R_v/V_v$.  Combining these results leads to
\eq{cooling2}{\bar q(t) \propto 
\frac{3-\eta}{\eta} f_b \left(\frac{3-\eta}{4\pi G}
\frac{f_b}{f(T)}\right)^{(3-\eta)/\eta}\left(G\rho_v(t)\right)^{3/2\eta}\ .}
For shorthand, we have written
\eq{cooling3}{f(T) = \frac{3}{2}\frac{kT m_H^2}{\mu X^2\Omega(T)}\ .}
From \cite{Sutherland}, the cooling function is known to be
\eq{cooling4}{\Omega(10^7\textnormal{K}) = 10^{-23}\, \textnormal{erg}\cdot
\textnormal{cm}^3\cdot \textnormal{s}^{-1}}
with little variation over several orders of magnitude in the temperature.

It is not, however, necessary for our purposes to compute the cooling and
star formation rates to this level of detail.  The cooling rate 
(\ref{cooling2}) depends on
time only through the virial density $\rho_v$, which is, to within an order
of magnitude, well-approximated by $10M_P^2 H^2$.  Therefore, it is reasonable
to fit the star formation rate (\ref{cooling1}) by a function of $H$, which
HS give as
\eq{sf2}{q(t)\propto H^{3/\eta} \frac{\bar H^{3/\eta}}{(H^{2m/3}+
\bar H^{2m/3})^{9/2m\eta}}\ ,}
with $m\sim 6$ and $\bar H \sim 10 H_0$ providing good fits to simulations.
This can be rewritten in the form (\ref{qt}).
Again, we hope to correct this inaccuracy in future work.

\subsection{Synthesis}
The final expression for the HS star formation rate can be simplified
by the approximation in terms of elementary functions 
of the error function which appears in the
Press-Schechter theory,
\beq
\mathrm{erf}(y) \approx  1 - \frac{1}{1 + \sqrt{\pi}y e^{y^2}}\ .
\eeq
Using constants $\delta_c = 1.6868$ and $a = 0.707$,\footnote{The constant
$a$ is introduced to modify the Press-Schechter fraction along the lines
of the Sheth-Torman fraction.}
HS find that
\beq
\label{complicated}
\dot\rho_\star\! = q(t)\left( 1\!- 
\!\frac{1}{3} \mathrm{erf}\left(\sqrt{\frac{a}{2}}
\frac{\delta_c}{\sigma_4}\right) \!
-\!\frac{2}{3}\mathrm{erf}\left(\sqrt{\frac{a}{2}}
\frac{\delta_c}{\sigma_{6.5}}\right) 
\right) ,
\eeq
where $\sigma_n$ stands for fluctuations at the mass scale which
virializes at temperature $10^n$ K.  The relation between mass and 
temperature at time $t$ is found to be 
\beq
M = \frac{(2kT/\mu_m)^{3/2}}{\rho_v}\approx 
\frac{(2kT/\mu_m)^{3/2}}{10 G H(t)}\ ,
\eeq
where $k$ is Boltzmann's constant and $M$ depends on the Hubble
parameter at the given time. 
Using the measured value $\xi/M_p = 3.3\times
10^{-28}$ for the matter abundance, the dimensionless mass parameter
$\mu_n$ corresponding to a virialization temperature 
$T = 10^n$ K is found to be
\beq
\mu_n = 1.1\times 10^{3 + 1.5 n} \left(\frac{\mathrm{Gyr}}{H(t)}
\right)\ .
\eeq
Using (\ref{spectrum2},\ref{growthG}), the corresponding
fluctuation amplitude is given by
\beq
\sigma_n = Q \Sigma(\mu_n) D(t)\ .
\label{sigmaeq}
\eeq

A simpler version of the SFR is also suggested by HS, namely
\beq
\label{SFR_A}
\dot\rho_\star = q(t)\left( 1- \mathrm{erf}\left(\sqrt{\frac{a}{2}}\,
\frac{\delta_c}{\sigma_4}\right) 
\right)\ .
\eeq
Our numerical comparisons indicate that this differs very little from
the version (\ref{complicated}), so we adopt (\ref{SFR_A}) for our
subsequent analysis.  This is our final SFR, given earlier in (\ref{SFR}).

\end{document}